\documentclass[aps,prb,twocolumn,floats,showpacs,superscriptaddress]{revtex4-1}

\usepackage{graphicx,epsfig}% Include figure files
\usepackage{times}
\usepackage{graphics,dcolumn,bm,float}
\usepackage{amssymb,amsmath,rotate,color}
\usepackage[title,titletoc,toc]{appendix}
\usepackage{graphicx}
\usepackage{wrapfig}

\usepackage{lipsum}
\usepackage{mwe}

\usepackage[pagebackref=false,colorlinks,linkcolor=blue,citecolor=blue,urlcolor=magenta]{hyperref}

\usepackage[mathlines]{lineno}
\usepackage{hyperref}
\usepackage{breqn}
\usepackage{bbold}
\usepackage{pgfplots}
\usepackage{float}
\usepackage{tkz-euclide}
\usepackage{braket}
\usepackage{physics}
\usepackage[caption=false]{subfig}
\usepackage[export]{adjustbox}
\usepackage{tikz}
\usetikzlibrary{through,calc}
\usetikzlibrary{positioning}

\newcommand{\be}{\begin{equation}}
\newcommand{\ee}{\end{equation}}
\newcommand{\ep}{\epsilon}

\newcommand{\bearr}{\begin{eqnarray}}
\newcommand{\eearr}{\end{eqnarray}}
\newcommand{\nn}{\nonumber}

\newcommand{\bsp}{{\boldsymbol{p}}}
\newcommand{\bsq}{{\boldsymbol{q}}}
\newcommand{\bsk}{{\boldsymbol{k}}}

\newcommand{\bs}{\boldsymbol}

\begin{document}
\preprint{}
\title{Kinky plasmons in double layers of borophene-borophene and borophene-graphene}
\author{Z. Jalali-Mola}
\email{jalali@physics.sharif.edu}
\affiliation{
Department of Physics$,$ Sharif University of  Technology$,$ Tehran 11155-9161$,$ Iran
}

\author{S.A. Jafari}
\email{jafari@sharif.edu}
\affiliation{
Department of Physics$,$ Sharif University of  Technology$,$ Tehran 11155-9161$,$ Iran
}
\affiliation{Center of excellence for Complex Systems and Condensed Matter (CSCM)$,$ Sharif University of Technology$,$ Tehran 1458889694$,$ Iran}
%\affiliation{Theoretische Physik, Universit\"at Duisburg-Essen, D-47048 Duisburg, Germany}

\date{\today}

\begin{abstract}
We investigate the collective plasmon modes in double layer of two dimensional materials where either one or both of the layers have tilted Dirac cone.
Consistent with quite generic hydrodynamic treatment, similar to double layer graphene systems we find two branches of plasmons.
The in-phase oscillations of the two layers disperses as $\sqrt q$, while the out-of-phase mode disperses as $q$. When even one of the layers hosts
tilted Dirac cone spectrum, the plasmonic kink which is a salient feature of a monolayer of tilted Dirac cone is inherited by both of these branches. In
double layers composed of graphene (nontilted) and borophene (tilted) where the two layers have two different Fermi velocities, the velocity
scale of plasmonic modes is set by the greater of the two. 
The kink always takes place when each plasmon mode crosses an energy scale $\omega_{\rm kink}$. When the two layers
have different chemical potentials, there will be two such scales, and each of in-phase and out-of-phase modes develop two kinks.
Moreover we find that an additional linearly dispersing overdamped mode of monolayer tilted Dirac cone system survives in the 
double layer system.
\end{abstract}



\maketitle
\narrowtext

\section{Introduction}
The concept of two dimensional (2D) Dirac materials began with graphene~\cite{Novoselov2004,castro2012,Novoselov1379,Novoselov2012,Novoselov2004,castro}
which was not only a big step in entering the 2D world but also a playground for 2+1 dimensional quantum field theories. When the Dirac theory
comes to condensed matter, it can be deformed in many ways~\cite{Vozmediano2016}. One interesting deformation of the Dirac cone is to tilt it~\cite{Cabra2013}.
There are already materials which host tilted 2+1 dimensional Dirac cones. In addition to organic conductor 
$\alpha$-BEDT~\cite{Tajima2009,Kobayashi2009,Katayama2006,Goerbig2008,Hirata2016} which has a weak coupling between layers,
a recent example of monolayer borophene~\cite{Mannix15,Cheng2016,Feng2017} has also been added to the list of tilted Dirac cone materials. 
The effect of tilting and anisotropic Fermi velocity on the electronic and collective properties of tilted Dirac cone
has been investigated~\cite{Nishine2010,Nishine2011,Goerbig2014,Suzumura2015,Tarun2017,Tilted2018}. 
In Ref.~\onlinecite{Tilted2018}, we have obtained an analytic representation of the polarization function for
tilted Dirac cone systems from which we have found a kink in the plasmon dispersion.
Moreover strong enough tilt gives rise to an additional overdamped plasmon mode which energetically lives in the intraband
particle-hole continuum (PHC). The undamped and overdamped plasmon modes at long wavelengths limit have square 
root and linear dispersion and  both of them are depended to the direction of the momentum $\bsq$. 

Not only the physics of a 2D monolayer borophene as a prototypical tilted Dirac cone material is interesting,
but it also possible to think of multilayer of such 2D systems composed of tilted Dirac cones, or combination
of tilted and upright Dirac cones. When these layers come close together in a periodic arrangement,
situations with potentially new physics can be created. The early investigations of the heterostructure of 2D systems, especially 2D 
electron gas dates back to the development of molecular beam epitaxy, which was 
employed to explore macroscopic sample of A-B super lattice like Ga and As compound.~\cite{Dingle1980}
The simplest of such systems are double layers~\cite{Jafari2014TI}.
When the layers are far enough to prevent their band overlap, the collective characteristic of the heterostructure of 2D systems will be different 
from monolayer one. The double layer combination of 2D electron gas in a either uniform dielectric 
background~\cite{Chang1980,DasSarma1981,Das-sarma1982,Olego1982,Pinczuk1986,Santoro1988,Hwang2009,Stauber2012} or arbitrary dielectric
media~\cite{Profumo2010,Profumo2012,Gan2012} forms two plasmon modes
corresponding to in-phase and out-of-phase plasmon oscillations of individual layers.  
The in-phase mode which appears in higher energy, in the long wave length limit disperses as $\omega\sim\sqrt{q}$,
while the out-of-phase mode disperses linearly $\omega\sim q$.
Indeed the square root behavior of a monolayer follows from a general hydrodynamic treatment, and is independent of 
macroscopic details~\cite{Fetter1973}, and holds for every 2D electron gas. 

Given that in addition to organic materials a 2D monolayer of borophene hosts tilted Dirac cone, it is timely to consider
the double layers of tilted Dirac cone systems. This can be either the double layer of borophene-borophene (DLB) or a
double layer of borophene-graphene (DLBG) where the borophene hosts tilted Dirac cone spectrum, while graphene
hosts upright Dirac cone. Based on our analytical calculation of the polarization function for tilted Dirac cone~\cite{Tilted2018},
we will investigate how the kink feature of monolayer tilted Dirac cone shows up in the double layer setting.
In the double layer system, the PHC will be union of the PHC of individual layers. We have
established (and will further establish) that in monolayer tilted Dirac cone system, there exists a $\omega_{\rm kink}(\bsq,\eta)$ curve
at which a kink in the plasmon dispersion appears which is controlled by the tilt parameter $\eta$. 
In this work, we find that in DLB systems when the dopings in two layers are different, there will be two such
scales, and therefore the number of kinks in each of the in-phase and out-of-phase modes will be doubled. 
More interestingly in the DLBG system, although the graphene layer does not have a kink in the decoupled limit,
as a result of coupling by Coulomb forces to the borophene layer, both resulting plasmon modes will develop
a kink at the $\omega_{\rm kink}$ energy scale. 
Another feature of monolayer tilted Dirac cone system is the presence of an additional plasmon mode
which disperses linearly and is heavily damped. We show that in DLB system there is only one such mode,
implying that the in-phase overdamped plasmons survive in DLB, while the out-of-phase mode does not exist. 

This paper is organized as follows. In section~\ref{model} we give a brief introduction double layer dielectric function and 
represent the tilted Dirac cone Hamiltonian. Then we analytically study the plasmon modes in the long wavelength limit. In section~\ref{DLB} we 
investigate the role of distance, tilting and doping in DLB.
In section~\ref{DLBG}, the plasmon modes are studied in double layer of borophene-graphene.  
Section~\ref{ODP} deals with the additional overdamped plasmon branch. 
The paper ends with a summary in section~\ref{summary}.

\section{Formulation of plasmons in DLB}\label{model}
To describe the collective excitations of the borophene layers in our double layer system, we need the linear response function of charge density to the external 
potential and the dielectric function for the double layer system. To begin, the Hamiltonian for DLB which are separated along $z$ direction
and are interacting via the long range Coulomb interacton~\cite{Das-sarma1982,Hwang2009}
\bearr
H=&&\sum_{l,\sigma,\bsk}\psi^{\dag}_{\bsk,\sigma,l} h(\bsk) \psi_{\bsk,\sigma,l}+\\
&&\frac{1}{2A}\sum_{l,l',\sigma,\sigma',\bsp,\bsq,\bsk} V_{ll'}(\bsq) 
   \psi^{\dag}_{\bsp+\bsq,\sigma,l}\psi^{\dag}_{\bsk-\bsq,\sigma',l'}\psi_{\bsk,\sigma',l'}\psi_{\bsp,\sigma,l},\nn
\label{total-hamiltonian}
\eearr 
where, $h(\bsk)$ is the Hamiltonian for tilted Dirac cone, $l,l'=1,2$ are layer indices, and $V_{ll'}$ denotes the 
interlayer ($l\neq l'$) and intralayer ($l=l'$) Coulomb matrix element. 
The operator $\psi_{\bsp,\sigma,l}$ anihilates an electron at Bloch state $\bsp$, with $\sigma$ in layer $l$ of area $A$.
The tilted Dirac Hamiltonian which describes the low energy electronic properties of borophene and organic conductors (under high pressure)
is given by~\cite{Goerbig2014,Proskurin2014,Tajima2009},
 \begin{equation}
h^{R/L}(\bsk)=\pm\hbar \begin{pmatrix}  v_{x0}k_x + v_{y0}k_y &  v_x k_x\mp i v_y k_y\\    v_x k_x\pm i v_y k_y &  v_{x0}k_x + v_{y0}k_y \end{pmatrix}.
  \label{matrixform}
 \end{equation}
 Here, L (R) stands for left (right) valley, the diagonal Fermi velocities $v_{x0}, v_{y0}$ represent the tilt of the Dirac cone and the 
off-diagonal Fermi velocities $v_x\neq v_y$ represent the anisotropic of the principal Fermi velocity. 
As a special case of Eq.~\eqref{matrixform}, if we assume that the diagonal Fermi velocity is zero and off-diagonal Fermi velocities are equal (symmetric), 
it will reduce to the graphene case. The noninteracting single-particle energy and eigenstates of tilted Dirac fermions after affecting a transformation on Cartesian coordinate $k_x, k_y $,
are given by
 \begin{align}
 &E^{R/L}_{\lambda}(\bsk)=\pm \hbar v_x (\eta k_x +\lambda k),\nn\\
   &\ket{\bsk,\lambda}^{R}= \frac{1}{\sqrt{2}}\begin{pmatrix} 1 \\ \lambda e^{i\theta_k} \end{pmatrix}
   ,~~~\ket{\bsk,\lambda}^{L}=\frac{1}{\sqrt{2}} \begin{pmatrix} 1 \\ -\lambda e^{-i\theta_k} \end{pmatrix},
  \label{energydispersion}
 \end{align}
where $\lambda=\pm$ and the tilt parameter $\eta$ is defined as
\be
 \eta=\sqrt{\frac{v_{x0}^2}{v_x^2}+\frac{v_{y0}^2}{v_y^2}}.
 \label{eta} 
\ee

The intralayer and interlayer Coulomb interaction for the pair parallel of 2D borophene layer in a medium with different dielectric constant 
around ($\ep_1, \ep_d, \ep_2$ form top to bottom) is given by\cite{Profumo2010}
\bearr
&&V_{11}=\frac{4 \pi e^2}{q D(q)}\bigg((\ep_d+\ep_2) e^{qd}+(\ep_d-\ep_2)e^{-qd}\bigg),\nn\\&&
V_{21}=V_{12}=\frac{8 \pi e^2}{q D(q)}(\ep_d),
\eearr
where,
\be
D(q)=(\ep_1+\ep_d)(\ep_d+\ep_2)e^{qd}+(\ep_1-\ep_d)(\ep_d-\ep_2)e^{-qd}.
\ee
Here, it has been assumed that the first (second) layer, $l=1$ ($l=2$), is located at $z=d$ ($0$) 
and has been sandwiched between two dielectric media $\ep_1$ at the top and $\ep_d$ ($\ep_d$ and $\ep_2$ at the bottom). 
Hence, $V_{22}$ can be find thorough $V_{11}$ by changing variable $\ep_2\to \ep_1$. 

 Within the random phase approximation (RPA), and ignoring the interlayer PH propagators, 
 the dressed density response function can be expressed as~\cite{DasSarma1981,Hwang2009,Das-sarma1982,Santoro1988}
 \be
 \chi(\bsq,\omega)=\chi_0(\bsq,\omega)\mathbb{1}-\bs{V}(q),
 \label{dressed}
 \ee
with  $\bs{V}(q)$ as electron-electron interaction matrix (in the space of layer indices) and $\chi_0$ as noninteracting density response function in which as long as the band overlap between layers is absent, the off diagonal (interlayer) density response tensor element is zero, and therefor a unit matrix on the right hand side
multiplies the scalar $\chi_0$. The dielectric function derived from Eq.~\eqref{dressed} is given by the matrix equation
\be
\bs{\varepsilon}(\bsq,\omega)=\mathbb{1}-\chi_0(\bsq,\omega) \bs{V}(q).
\label{dielectric.eqn}
\ee
Now the dispersion of collective excitations is obtained by 
\be
  \det\bs{\varepsilon}(\bsq,\omega)=0.
  \label{secular.eqn}
\ee

To begin investigating the plasmon mode properties in the DLB we first analyze the density response and its plasmons in long wavelength limit, analytically.
 Using linear response theory, the density fluctuation of the noninteracting borophene monolayer in the presence of external electromagnetic filed is given by,
\bearr
&& \chi_0(\boldsymbol{q},\omega)=\\&&   \frac{g  \gamma^2}{A }  \sum_{\bsk,\lambda , \lambda'=\pm} \frac{n_{\bsk,\lambda}-n_{\bsk',\lambda'}}
{\hbar \omega+E_{\bsk,\lambda}-E_{\bsk',\lambda'}+i0^+} f_{\lambda, \lambda'} (\bsk, \bsk')\nn,
\label{pai}
 \eearr
where $\bsk'=\bsk+\bsq$, $n_{\bsk,\lambda}$ is Fermi distribution function with $\lambda$ as band index, the factor $\gamma^2=v_x/v_y$ is Jacobian transformation,
$g$ includes spin degeneracy, $A$ is area of system and the form factor $f_{\lambda, \lambda'} (\bsk, \bsk')$ is defined as an expectation value of 
the density operator $\sigma_0=\mathbb{1}$ between two eigenstates $\ket{\bsk,\lambda}$, $\ket{\bsk',\lambda'}$ of the tilted Dirac cone, which is given by
\be
f_{\lambda, \lambda'} (\bsk, \bsk')=\bra{\bsk,\lambda}\ket{\bsk',\lambda'}= \frac{1}{2} (1+\lambda \lambda'\cos(\theta_\bsk-\theta_{\bsk'})),
\ee
where $\theta_{\bsk}$ is a direction of  momentum $\bsk$ to $x$ axis. 

Here we just consider contribution of one valley (say right) in each layer and ignore the effect of intervalley process, which
require large momentum transfer. The analytical result of noninteracting density response function has been calculated in 
Ref.~\onlinecite{Tilted2018} by the present authors. In the long wavelength limit $q \to 0$ this result can be summarized as,
\bearr
 \chi_0(q\rightarrow0,\omega)\approx \begin{cases}
     \frac{\mu q^2}{4\pi \omega^2} F(\eta) &\eta\ll q ~~ ,~~ \frac{\omega \eta }{q}\ll 1,     \\
    
     \frac{\mu q^2 }{4 \pi\omega^2 }G(\eta,\phi) & 
          \eta\gg q ~~,~~ \frac{\omega \eta }{q}\gg 1, 
 \label{longwl.eqn}
     \end{cases}
\eearr
where,
\bearr
&&G(\eta,\phi)=\frac{g}{4\pi \hbar^2 \eta^2 v_x v_y}\bigg(\cos2\phi+\frac{\eta^2+(\eta^2-2)\cos2\phi}{\sqrt{1-\eta^2}}\bigg),
\nn\\&&
F(\eta)=\frac{g}{4\pi \hbar^2 v_x v_y} (1-\frac{2\omega\eta}{q}).
\label{DF.eqn}
\eearr

It can be easily seen that the density response function in Eq.~\eqref{DF.eqn} depends on the tilting parameter $\eta$, and the direction $\phi$ of wave vector 
$\bsq$. Note that in the limit $\eta\to 0$, the ($\eta\omega/q\ll1$ piece of the) above density response function reduces to graphene. Furthermore, 
the collective excitations of monolayer borophene is a function of $\eta$ and $\bsq$ and show square root behavior as typical plasmon in 
2D material~\cite{Fetter1973,Nishine2010,Nishine2011,Tilted2018}. However, in the DLB layer, the collective modes are different from monolayer. 
For quite general values of $\ep_1,\ep_d,\ep_d$, there will be two branches of plasmons for DLB system, one dispersing as 
$q^{1/2}$ and the other dispersing as $q^1$. For simplicity let us assume that $\ep_1=\ep_d=\ep_2$ are a uniform background
dielectric constant. Combining the above equations with Eq.~\eqref{dielectric.eqn} to solve the secular equation~\eqref{secular.eqn} gives
\bearr
&&\omega_+^2\approx \begin{cases}
     e^2 q(\mu_1+\mu_2)  F(\eta) &  \eta\ll q ~~ ,~~ \frac{\omega \eta }{q}\ll 1,  \\
     
     e^2 q (\mu_1+\mu_2)G(\eta,\phi) &  \eta\gg q ~~,~~ \frac{\omega \eta }{q}\gg 1, 
 \label{collectivep.eqn}
     \end{cases}
\eearr
and,
\bearr
&&\omega_-^2\approx \begin{cases}
     2e^2 q^2 d F(\eta)/(\mu_1+\mu_2) &  \eta\ll q ~~ ,~~ \frac{\omega \eta }{q}\ll 1, \\
     
      2e^2 q^2 d G(\eta,\phi)/(\mu_1+\mu_2) &\eta\gg q ~~,~~ \frac{\omega \eta }{q}\gg 1,
 \label{collectiven.eqn}
     \end{cases}
\eearr
The $\omega_+\propto \sqrt q$ is the in-phase oscillations of charge density in two layers,
and therefore conforms to the generic $\sqrt q$ hydrodynamic behavior of 2D systems~\cite{Fetter1973}.
The $\omega_-$ is the out-of-phase collective oscillations of the density in two layers. 
Some times the $\omega_-$ is referred to as the "acoustic" plasmon which is misleading.
The acoustic modes in phonon systems refer to the in-phase oscillations of the ions in the
same unit cell, while here the linearly dispersing plasmon mode corresponds to out-of-phase oscillations. 
It is interesting to note that the in-phase plasmon mode does not depend on separation $d$ of the two
layers, while the out-of-phase (linearly dispersing) is proportional to $\sqrt{d}$, and its energy increases by
increasing the separation $d$ of the layers. 

Both plasmonic branches depend on the chemical potential of each layer and the linear one is also sensitively dependent on the separation of the layers. 
Note that up to this point we have assumed that except for the chemical potential, all other parameters of the two layers forming the DLB are the same.
In addition, if we consider different dielectric media, the qualitative behavior remains the same, but the two plasmonic branches will disperse at lower 
energy and  the group velocity of acoustic mode will be modified~\cite{Profumo2012}. In this paper we assume that our double layer system is placed in the 
medium with a uniform background dielectric constant, i.e. $\ep_1=\ep_d=\ep_2$. Moreover, when studying bilayers composed of graphene and borophene,
their Fermi velocities is assumed to be $v_F=c/300, c/1000$, respectively, where $c$ is the light velocity. 
The fine structure constant is $e^2/\hbar c=1/137$. 
We also assume that the Fermi velocity in $x$ and $y$ directions are the same, $v_x=v_y=v_{F}$.
Moreover, the kink feature in which we are interested is best seen for the direction $\phi=\pi/2$ of the momentum $\bsq$. 
So we will report our plots for this direction. 
energy $\hbar\omega/\mu_1$ (momentum $q/k_{F_1} $) where the subscript $1$ refers to layer $1$ which is 
taken as reference in case the corresponding quantities are different from layer $2$, and the vector
$\bsq$ is along the $y$ direction. 

%\textcolor{blue}{In the opposite limit of
%weak coupling situation ($q a\gg 1$), one recovers N
%independent plasmons of monolayer layer,. we consider $v_b=3\time10^5$ }

\section{Borophene-borophene double layer}\label{DLB}
In this section we consider, the pair of parallel borophene layers which are placed in background dielectric constant ($\ep_1=\ep_d=\ep_2=1$) 
and separated by a distance $d$ in $z$ direction and investigate the dependence of the two plasmon dispersions on
various parameters such as the tilting parameters ($\eta_1,\eta_2$), chemical potentials ($\mu_1,\mu_2$).
The background dielectric constant appears as an overall constant that reduces $V(\bsq)$
to which the results are not very sensitive. So we take the background dielectric constant to be $1$.
Moreover, all plots will be in the $(\omega,\bsq)$ space, where the vertical (horizontal) axis is the dimensionless 

\subsection{Distance and tilt dependence}
\begin{figure}[t]
 \includegraphics[width = .47\textwidth] {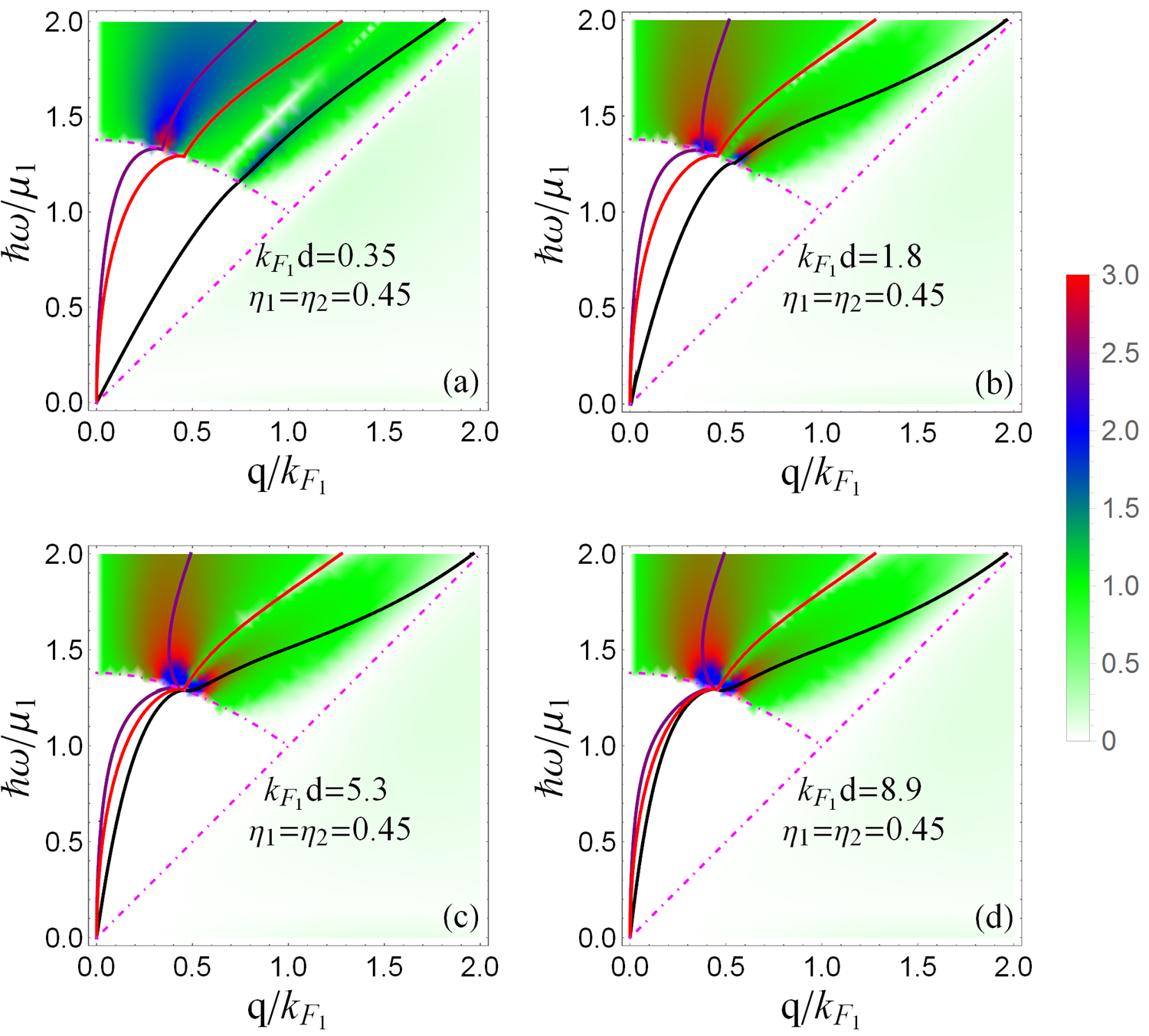}
   \caption{(Color online)  Plasmon mode dispersions in DLB system with the same chemical potential ($\mu_1=\mu_2$) and same tilting parameter ($\eta_1=\eta_2=0.45$) 
accompanied with intensity plot of loss function $|\Im \varepsilon^{-1}(\bsq,\omega)|$. Here, the vertical (horizontal) axis is the dimensionless 
energy $\hbar\omega/\mu_1$ (momentum $q/k_{F_1} $) and the direction of $\bsq$ is fixed by $\phi=\pi/2$. The spacing between layers, in various panels 
are: (a) $k_{F_1} d=.35$, (b) $1.8$, (c) $5.3$, and (d) $8.9$. Purple, black and red solid lines are the in-phase ($\omega_+$), out-of-phase ($\omega_-$) and 
monolayer plasmon modes, respectively. The pink  dotdashed line and curve are the borders of  intra-band ($\omega_{\rm kink}$) and inter-band ($\omega_s$) PHC, respectively. 
See the text for details.} 
 \label{dif-dis.fig}
\end{figure}

To begin with, we consider DLB with both layers at the same chemical potential ($\mu_1=\mu_2$).
Since both layers are made of borophene, they have the same tilting parameter. We take the tilt parameter to be 
$\eta_1=\eta_2=0.45$~\cite{Goerbig2014}. 
In Fig.~\ref{dif-dis.fig}, we show how the plasmon modes disperse by increasing their separation in $z$ direction. 
Here, we have plotted the plasmon mode dispersion along with the loss function $|\Im \varepsilon^{-1}(\bsq,\omega)|$ to clearly show the damping structure. 
We have assumed in all panels of Fig.~\ref{dif-dis.fig} that the direction of
$\bsq$ is fixed by $\phi=\pi/2$. 
The separation between the two borophene layers in each panel is: (a) $k_{F_1}d=0.35$, (b) $k_{F_1}d=1.8$, (c) $k_{F_1} d=5.3$ and (d) $k_{F_1} d=8.9$. 
In this figure, the in-phase (out-of-phase) plasmon mode i.e. $\omega_+$ ($\omega_-$) has been shown with purple (black) curves. 
The red line denotes the plasmon mode for the monolayer borophene. When the separation becomes infinitely large, the two layers are expected to 
be decoupled, and therefore the in-phase and out-of-phase modes both become degenerate with the monolayer (red) mode. 
The boundary of  interband and intraband PHC which is defined by $\omega_{\rm kink}$ and $\omega_s$ has been shown with 
dotdashed curve and line, respectively. These boundaries are given by~\cite{Nishine2010,Tilted2018}
\begin{align}
   &\omega_{\rm kink}(\bsq,\eta)=v_Fq\eta\cos\phi+\frac{2\mu}{1-\eta^2}\\
   &-\sqrt{(v_Fq)^2+\frac{4 v_Fq \mu\eta \cos\phi }{1-\eta^2}+\left(\frac{2\mu\eta}{1-\eta^2}\right)^2}\nn\\
   \label{omega_kink.eqn}
   &\omega_s=v_Fq (1+\eta \cos\phi)
\end{align}
The reason the lower boundary of inter-band PHC is given the name $\omega_{\rm kink}$ is that in 
Ref.~\onlinecite{Tilted2018} we established that at this boundary the plasmon dispersion of a single-layer
tilted Dirac cone system develops a kink. 

As can be clearly seen in Fig.~\ref{dif-dis.fig}, both in-phase and out-of-phase (linear) plasmon mode in all panels
maintains their kink in the double layer system as well. Note that since tilt parameter for both layers is the
same, they are both characterized with the same $\omega_{\rm kink}$ curve. That is why the combined system develops 
its kinks on the same curve. Note that the in-phase mode is always above the single-layer mode, while the 
out-of-phase mode is always below the single-layer mode. This is consistent with picture of two harmonic
oscillators coupled via inter-layer Coulomb forces which then splits the degenerate modes into two,
lying above and below the degeneracy limit. Since the coupling between the layers becomes zero in the $d\to\infty$
limit, both curves must tend to the single-layer curve by increasing $d$. This can be clearly seen
by looking at the trends from panels (a) to (d). 
It is very pleasant to notice the distance dependence of linear mode. By increasing the distance of layers, 
the linear (out-of-phase) mode increases. This is consistent with our analytic result in Eq.~\eqref{collectiven.eqn}
which suggests that the linear mode depends on distance as $\sqrt d$. 
Upon entering the interband PHC, both modes acquire damping. The undamped portion of the dispersion 
relation conforms to intuition and both modes tend to the same monolayer dispersion when the distance
becomes very large. However the damped portion of the plasmon dispersions which are inside the interband PHC
do not degenerate to the same curve. This feature is similar to the case of double layer graphene~\cite{Hwang2009},
which is the special case where $\eta_1=\eta_2=0$. 

\begin{figure}[t]
 \includegraphics[width = .47\textwidth] {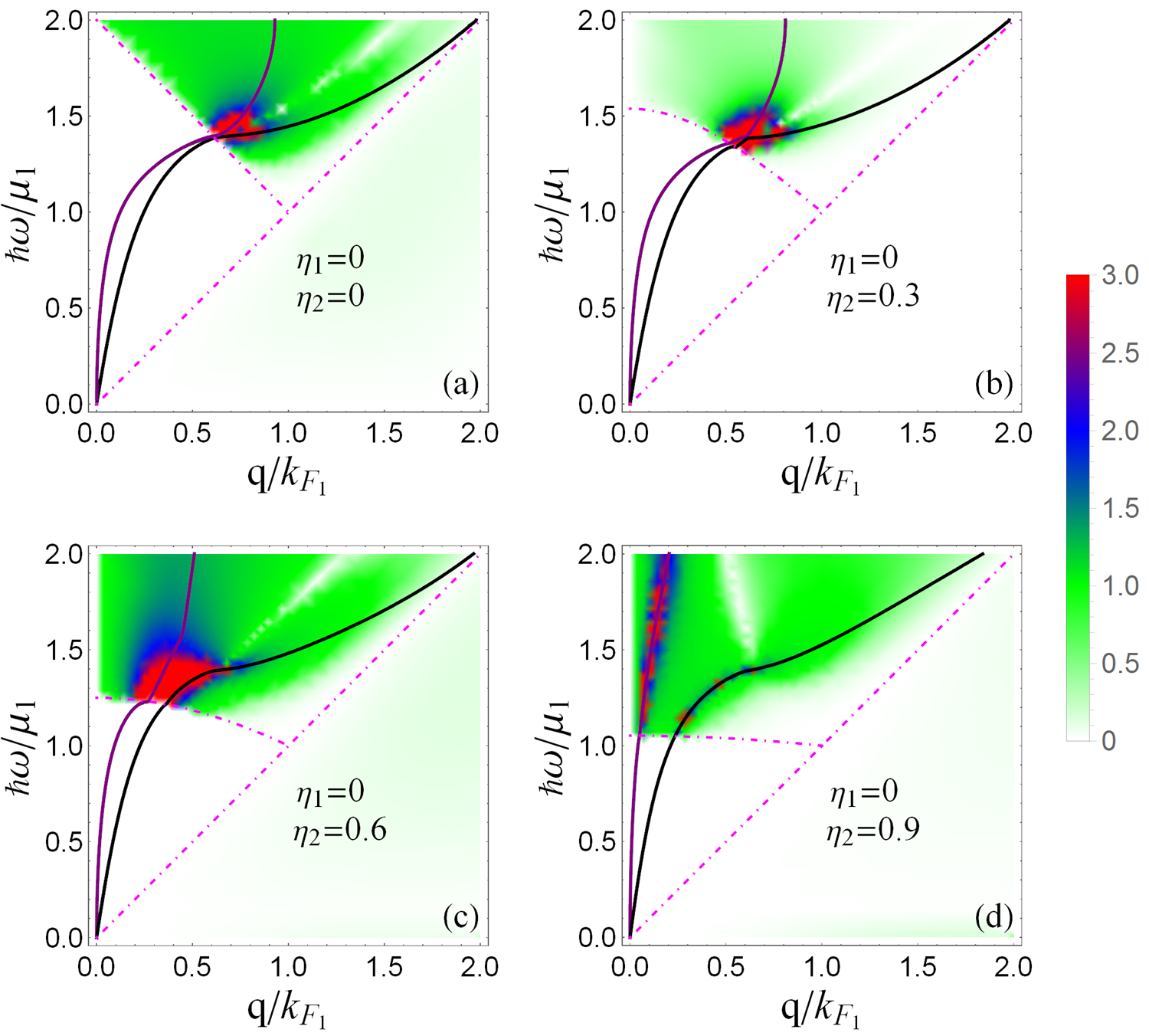}
   \caption{(Color online) Intensity plot of loss function $|\Im \varepsilon(\bsq,\omega)|$ and the plasmon dispersions for combination of  
two tilted Dirac cone layers. The tilt in the first layer is assumed to be zero, i.e. $\eta_1=0$, and we vary the tilt strength in the second
layer. The direction of $\bsq$ is fixed by $\phi=\pi/2$ and the layer spacing is $k_{F_1}d=5.3$. 
The vertical (horizontal) axis is the dimensionless energy $\omega/\mu_1$ (momentum $q/k_{F_1} $). 
The tilt parameter $\eta_2$ of the second layers is (a) $0$, (b) $0.3$, (c) $0.6$ and (d) $0.9$. 
Black and purple solid lines are the plasmon modes for $\omega_+$ and $\omega_-$, respectively. 
The dotdashed curves are the boundary of PHC as before.} 
 \label{eta0-eta.fig}
\end{figure}
To further investigate the role of tilt parameter, let us assume that one of the layers is not tilted, i.e. $\eta_1=0$,
and vary the tilt strength $\eta_2$ of the other layer. This will teach us how the kink which is the
hallmark of tilted Dirac cone evolves in a double layer system. For this purpose in Fig.~\ref{eta0-eta.fig}
we plot the plasmon dispersion for $\bsq$ in $\phi=\pi/2$ direction. The velocity of both layers are
assumed to be identical to that of borophene $\sim c/1000$. In all panels the distance is given by $k_{F_{1}}d=5.3$. 
The chemical potentials of both layers are also assumed to 
be the same, so that we only focus on the variation of $\eta_2$, as indicated in the legend of each panel. 
As in the Fig.~\ref{dif-dis.fig}, the in-phase and out-of-phase modes are plotted by purple and black lines and the 
PHC with dotdashed lines. 
The separation $d$ of the layers is chosen to be large enough such that the two modes in panel (a) are very
close to each other. As can be seen, the effect of tilt in the second layer is to push them away from 
each other. Furthermore, larger tilt in the second layer increases the energy of the in-phase mode. 

Now let us see how the kink is imparted to the two modes. 
As can be seen from panel (a) in Fig.~\ref{eta0-eta.fig} where both layers have zero tilting 
there is not kink whatsoever. By increasing $\eta_2=0.3,0.6,0.9$ as in panels (b), (c) and (d), both modes
develops a kink. This can be intuitively understood as follows: When the layers are decoupled, 
only the second layer has a kink, as $\eta_1=0$. But when they are coupled by Coulomb forces, the in-phase and out-of-phase
eigen-modes will be linear combinations of the modes in layers $1$ and $2$. Depending on the relative magnitudes
of the coefficients in the linear combination that forms the two eigen modes, the kink will be more manifest
in either or both of the symmetric and asymmetric modes. As can be seen in panel (b) for $\eta_2=0.3$, the kink
in the out-of-phase mode is more manifest, while in panel (c) corresponding to $\eta_2=0.6$, the kink
for the in-phase mode is more manifest. This observation can be analytically formalized as follows:
The eigen-modes for $\eta_1=0$ and $\eta_2\ne 0$ are given by
\begin{align}
 & \omega_+^2=\frac{e^2 q\mu }{2}  \big(F(\eta_1=0)+G(\eta_2,\phi)\big),\nn\\
  &\omega_-^2= e^2 q^2 \mu d \frac{F(\eta_1=0)G(\eta_2,\phi) }{F(\eta_1=0)+G(\eta_2,\phi) }.
 \label{collective-eta0-eta.eqn}
\end{align}
This is obtained by plugging the long wavelength expression of Eq.~\eqref{longwl.eqn} in the
characteristic equation~\eqref{secular.eqn}. 
Note that although this equation is valid in the hydrodynamic limit where $q$ is very 
small and kinks appear at higher $q$, but still this equation shows how the
function $G(\eta_2,\phi)$ enters both $\omega_\pm$ modes. This
function encodes information about the kink which is now shared by both $\omega_\pm$ modes.

\begin{figure}[t]
 \includegraphics[width = .47\textwidth] {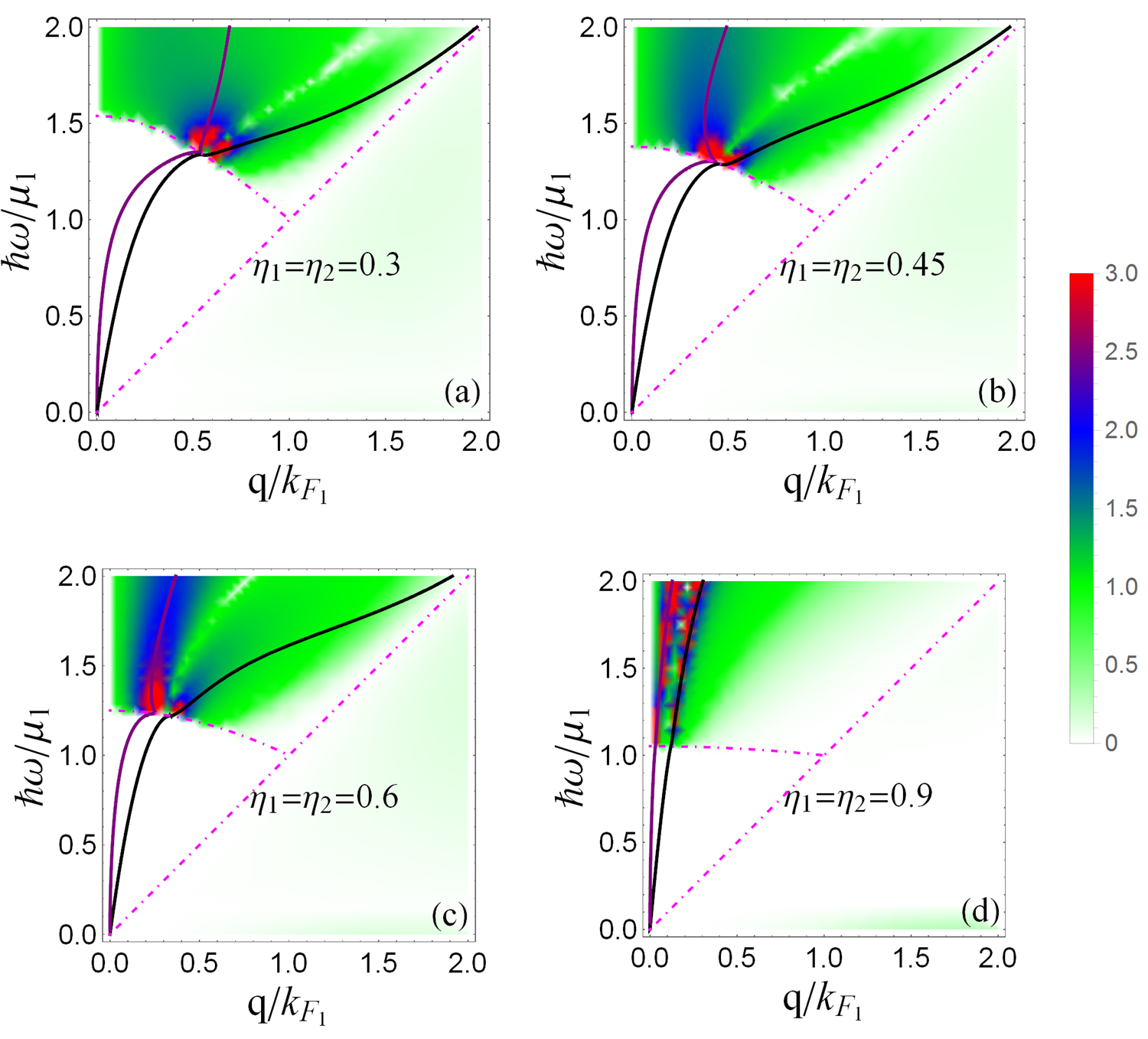}
   \caption{(Color online) The plasmon dispersions for DLB system with equal nonzero tilting parameter($\eta_1=\eta_2=\eta$) combined with the density plot of loss function $|\Im \varepsilon^{-1}(\bsq,\omega)|$. The direction of $\bsq$ is fixed by $\phi=\pi/2$ and the layer spacing is $k_{F_1}d=5.3$. The vertical (horizontal) axis is the dimensionless quantity $\omega/\mu_1$ ($q/k_{F_1} $) . The tilting parameter for panel (a), (b), (c), (d) is $\eta=0.3, 0.45, 0.6, 0.9$ respectively. Black and purple solid lines are the plasmon dispersions for $\omega_+$ and $\omega_-$, respectively. The dotted and Dashed green lines are the boundary of particle hole continuum. } 
 \label{eta-eta.fig}
\end{figure}

Now let us return to the problem of identically tilted layers. Again both layers have the same chemical potential ($\mu_1=\mu_2$), 
the same tilting parameter ($\eta_1=\eta_2$), and the same velocities. In Fig.~\ref{eta-eta.fig} we have plotted the symmetric and asymmetric plasmon modes for
different tilting parameter. In this figure the tilting parameter of both layers in panels (a), (b), (c) and (d) is given by 
$\eta_1=\eta_2=0.3, 0.45, 0.6, 0.9$, respectively. As before, the direction $\bsq$ is fixed at $y$ axis. 
It can be seen that by increasing the tilting parameter from panel (a) to (d), first of all 
both modes have kinks at $\omega_{\rm kink}$ energy scale. This is intuitive, as both layers
have their own kink at $\omega_{\rm kink}$, and so does their both symmetric and asymmetric combinations. 
Secondly by increasing the kink
the splitting between the modes on the $\omega_{\rm kink}$ boundary increases. Third, the 
dispersions become steeper by increasing the tilt strength. In particular note the very steep
dispersion in panel (d) which we have deliberately chosen plot for $\eta_1=\eta_2=0.9$. This
large group velocity can be understood from Eq.~\eqref{collectivep.eqn} and Eq.~\eqref{collectiven.eqn}.
Both these equations suggest that the plasmon energy depends on $\sqrt G$. On the other hand
according to Eq.~\eqref{DF.eqn}, at least near $\eta\sim 1$, the auxiliary function $G$ behaves as
\be
   G(\eta,\phi)\sim \frac{1-\cos 2\phi}{\sqrt{1-\eta^2}}.
\ee
This implies that for $\eta\approx 1^-$, both in-phase and out-of-phase modes behave like
\be
   \omega_{\pm}\sim \frac{\sin\phi}{(1-\eta^2)^{1/4}}
   \left\{\begin{matrix}
   q^{1/2}\\
   q^1
   \end{matrix}\right.
\ee
This singular behavior near $\eta\approx 1$ explains why the plasmon modes become steeper for very large tilting. 

\begin{figure}[t]
 \includegraphics[width = .47\textwidth] {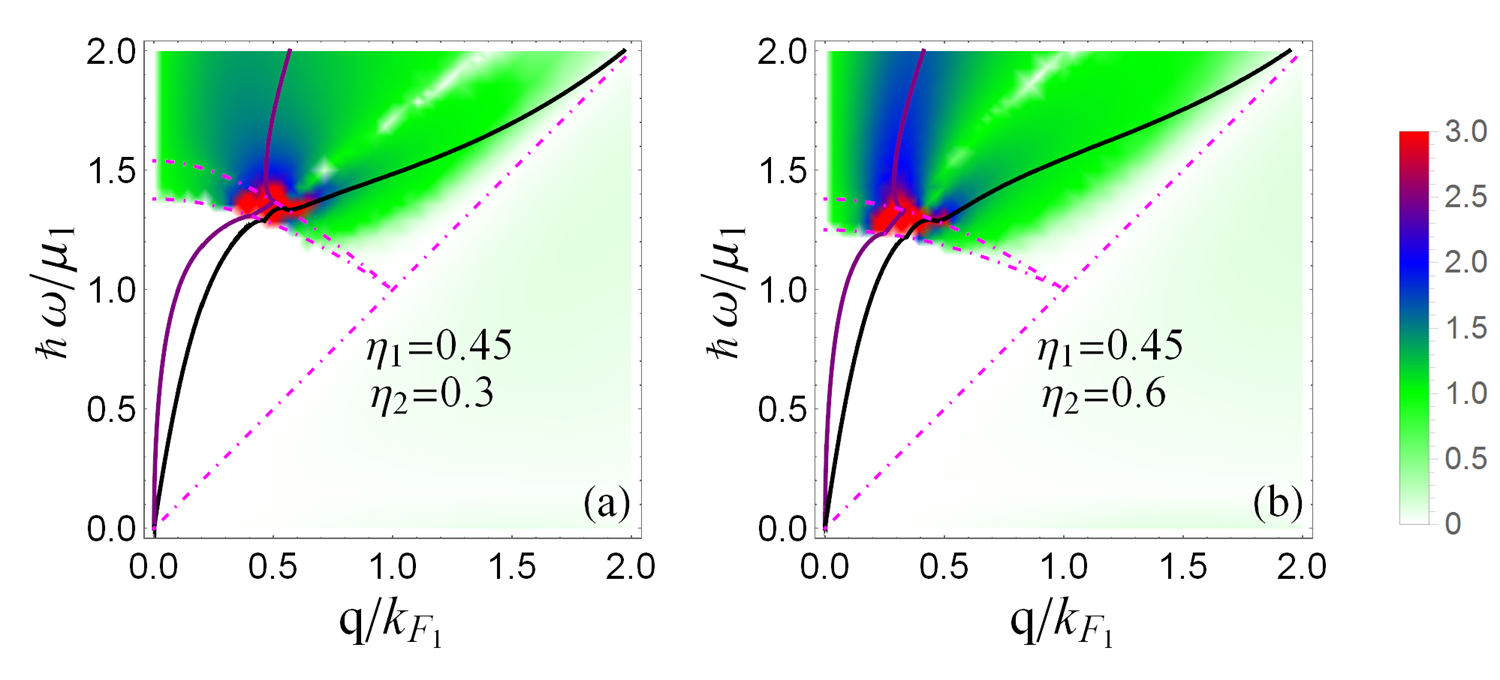}
   \caption{(Color online) Intensity plot of loss function $|\Im \varepsilon(\bsq,\omega)|$ and the plasmon dispersions for combination of  borophene layers with different nonzero tilting parameter. The direction of $\bsq$ is fixed by $\phi=\pi/2$ and the layer spacing is $k_{F_1}d=5.3$. The vertical (horizontal) axis is the dimensionless quantity $\omega/\mu_1$ ($q/k_{F_1} $) . The tilting of the first layer in all panels is $\eta_1=0.45$ and the other layer tilting in each panel is, $\eta_2=0.3$ in (a) and $0.6$ in (b). 
  Other conventions are the same as previous figures.} 
 \label{eta45-eta.fig}
\end{figure}

To establish the claim of our previous work~\cite{Tilted2018} that the kink is associated with the 
energy scale $\omega_{\rm kink}$ of Eq.~\eqref{omega_kink.eqn}, let us now introduce two such curves
corresponding to two different kinks. For this purpose in Fig.~\ref{eta45-eta.fig}, we 
consider DLB with different tilting parameter in each layer. We suppose the first layer has fixed tilting parameter $\eta_1=0.45$ and the 
other layer has a tilt parameter different from $\eta_1=0.45$. Panels (a) and (b) of this figure correspond to 
$\eta_2=0.3,0.6$, respectively. Since each $\eta$ according to Eq.~\eqref{omega_kink.eqn} gives rise to a
distinct boundary $\omega_{\rm kink}$ for the inter-band particle-hole excitations, with two different $\eta_1\ne\eta_2$
we will have two of them which are plotted as dotdashed lines in Fig.~\ref{eta45-eta.fig}. 
Here again the general trends of plasmon modes are the same as Fig.~\ref{eta0-eta.fig} and Fig.~\ref{eta-eta.fig}.
But an important difference is that here as a result of two different tilting parameters of layers, we have two different
boundaries given by $\omega_{\rm kink}(\eta_1)$ and $\omega_{\rm kink}(\eta_2)$. Now every one of the plasmon branches --
either in-phase or out-of-phase modes -- develops a kink upon crossing every one of these boundaries. This
gives us a total number of {\rm four kinks} in the plasmon dispersion, two for each mode. 
Again this can be seen analytically. The eigen modes for arbitrary and nonzero $\eta_1$ and $\eta_2$
are given by
\begin{align}
 & \omega_+^2=\frac{e^2 q\mu }{2}  \big(G(\eta_1,\phi)+G(\eta_2,\phi)\big),\nn\\
  &\omega_-^2= e^2 q^2 \mu d \frac{G(\eta_1,\phi)G(\eta_2,\phi) }{G(\eta_1,\phi)+G(\eta_2,\phi) }.
 \label{collective-eta1-eta2.eqn}
\end{align}
Every $G$ factor contains its own kink information, and therefore both $\omega_\pm$ modes will 
inherit two kinks, one from the $G$ function of each leyer.

\subsection{Role of doping}

\begin{figure}[t]
 \includegraphics[width = .47\textwidth] {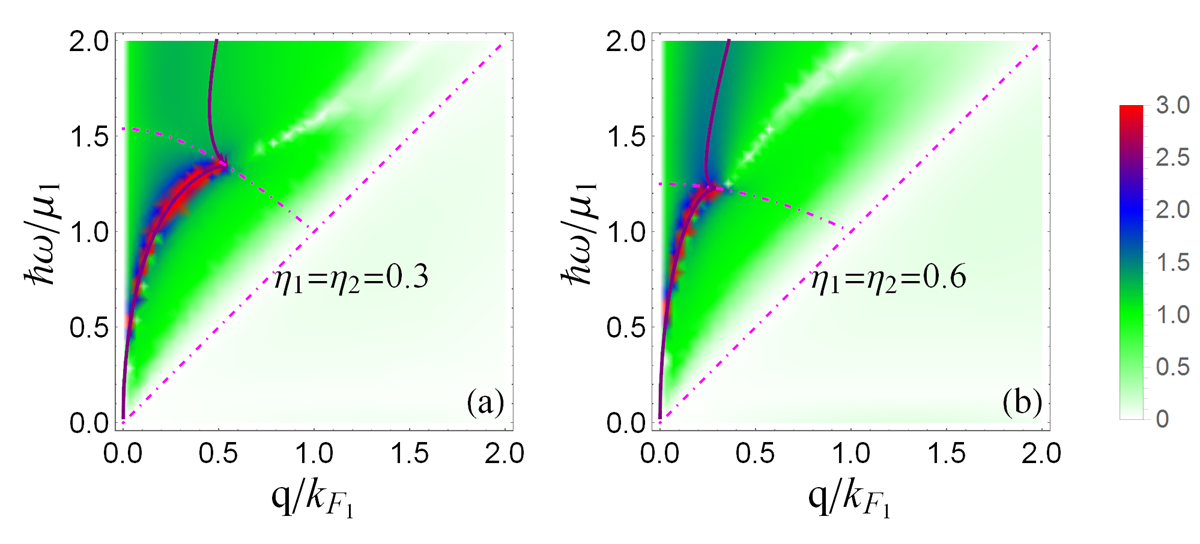}
   \caption{(Color online) The plasmon dispersions for DLB with undoped-doped combination of  borophene layer along with intensity 
plot of loss function $|\Im \varepsilon^{-1}(\bsq,\omega)|$ . The vertical (horizontal) axis is the dimensionless energy 
$\hbar \omega/\mu_1$ (momentum $q/k_{F_1} $). The direction $\phi$ of $\bsq$ in both panels is $\pi/2$.
The separation of layers is set by $k_{F_1}d=5.3$.
The tilting parameter in both layers is the same, $\eta_1=\eta_2$ which in panel (a) is set to $0.3$, 
while in panel (b) the tilt parameters are $0.6$. The purple curve is the in-phase. The out-of-phase mode is absent in this case.
The pink dotdashed lines are the boundary of interband and intraband PHC. Note that the $\omega_{\rm kink}$ boundary 
belongs only to the doped layer.} 
 \label{BB1.fig}
\end{figure}

An interesting lesseon can be learned by studying the plasmon modes of two borophene layers 
where layer $1$ is doped ($\mu_1\ne 0$), while the second layer is undoped ($\mu_2=0$). 
When such two layers are infinitely separated, such that the collective charge oscillations
in the two layers are decoupled, in layer $1$ we have standard plasmons, while in layer $2$, since
the doping level is zero, there are no plasmon oscillations at the RPA level~\cite{Triplet2017,mishchenko}.
Although there will be other types of spin-flip modes~\cite{Jafari2002,Jafari2004,Ebrahimkhas2009,Jafari2012,Ganesh2013,Jafari2014TI,Hedegard2015,Maslov2017}
Therefore in terms of counting the collective degrees of freedom, we only have one mode. 
When the two layers are brought closer at a distance of $k_{F_1}d=5.3$ as in Fig.~\ref{BB1.fig} 
to let them couple, it is not surprising to see that there is only one solution, which clearly
corresponds to the $\sqrt{q}$ dispersion. This is a further confirmation that the $\sqrt q$ mode
is indeed the in-phase mode. 

The PHC consist of two contributions. In the doped layer, there is a window below the $\omega_{\rm kink}$.
But in the undoped layer, this window is filled with interband PH excitations. Therefore the total PHC
which is the union of the PHC for the two layers, 
consists of no gapped (white) region which will then make the plasmon mode of essentially layer $1$ Landau
damped by creasing interband PH excitations in the layer $2$. 
Indeed we have checked that the dispersion of the present DLB system is almost degenerate with that
of a single layer $1$, as long as we are concerned with $\omega<\omega_{\rm kink}$.
However, for $\omega>\omega_{\rm kink}$ the plasmon branch enters
the interband PHC of layer $1$ itslef, and its dispersion is heavily affected by the presence of layer $2$, 
and it will no longer be nearly degenerate with the dispersion of a monolayer $1$.
Finally note that by increasing the common tilt parameter $\eta_1=\eta_2$ of the two layers, 
the energy of the plasmon mode increases. This is a generic behavior in all combinations,
as in e.g. Fig.~\ref{eta-eta.fig}.

\begin{figure}[b]
 \includegraphics[width = .45\textwidth] {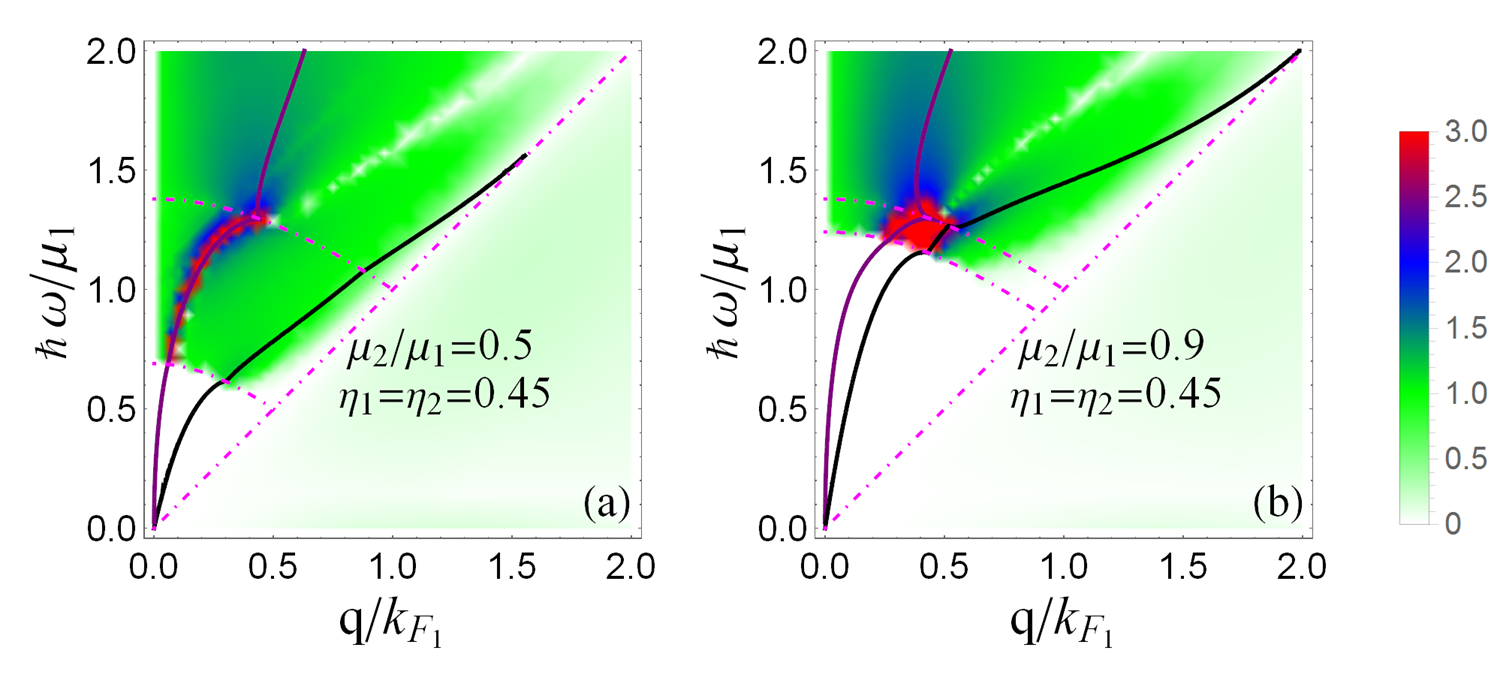}
   \caption{(Color online) The plasmon mode dispersions for DLB system with different chemical potential ($\mu_1\neq\mu_2)$ and same tilting parameter,
   $\eta_1=\eta_2=0.45$, and same Fermi velocity. The color code as before indicates the intensity plot of loss function $|\Im \varepsilon^{-1}(\bsq,\omega)|$. 
   The vertical (horizontal) axis is the dimensionless energy $\hbar\omega/\mu_1$ (momentum $q/k_{F_1} $).The separation of layers is set by $k_{F_1}d=5.3$. The direction of $\bsq$ is fixed in $y$ direction. 
   The doping ratios in panels (a) and (b) are $\mu_2/\mu_1=0.5$ and $\mu_2/\mu_1=0.9$, respectively. 
   Black and purple solid lines are the out-of phase and in-phse plasmon modes, respectively. 
   The pink dotdashed  lines are the boundary of interband and intraband borophene PHC. } 
 \label{BB2.fig}
\end{figure} 
Next, we assume in the DLB we have identically tilted layers with identical velocities, which are doped differently. The difference in doping is quantified by 
doping ratio. In Fig.~\ref{BB2.fig}, we have plotted the plasmon dispersion for DLB system, for the doping ratio  $\mu_2/\mu_1$ given by (a) $0.5$, and (b) $0.9$. 
The color code is the same as previous figures and  
the direction of wave vector $\bsq$ is fixed in $y$ direction. 
An important player in this case is the upper border $\omega_s$ of the intra-band PHC. 
As for the border $\omega_{\rm kink}$ of the interband PHC, there will be three possibilities
to form the interband particle-hole excitations: (i) within the layer $1$, (ii) within the layer $2$, (iii) cross layer involving
particle-hole excitations between layer $1$ and $2$. In the present approximation where interlayer PH propagators are 
not included, the third item above is absent. 
The lower bound $\omega_{\rm kink}$ of the intralayer interband for each of
the layers are plotted by dottdashed lines. As can be seen first of all, both modes when cross every one of these boundaries,
develop a kink. So we end up having two kinks for each mode. Second point to notice is that, in panel (b) where the chemical
potentials are closer to each other, the $\omega_{\rm kink}$ boundaries approach to each other. In this case, we have two
decent $\omega_\pm$ modes. However by reducing the ratio $\mu_2/\mu_1$, the nearly triangle shaped region shrinks more and more. 
As a result, the out-of-phase mode (black line) is attracted more and more to the intraband PHC. At the limit $\mu_2/\mu_1=0$
of Fig.~\ref{BB1.fig}, the out-of-phase mode is entirely swallowed by the intraband PHC. 

\section{Brophene-Graphene}\label{DLBG}
So far we have assumed that both layers are composed of borophene, such that the Fermi velocities are identical. 
In this section, we are going to study a double layer composed of borophene and graphene. In this case, a new
player will be the difference in the Fermi velocity of the two layers. The Fermi velocity sets the slope of 
the boundary $\omega_s$ of the intraband PHC for every layer. 

The monolayer graphene as a 2D Dirac material with Fermi velocity $c/300$ and borophene
layer as a 2D tilted Dirac material with Fermi velocity $\approx c/1000$ are a good candidate for constructing a double layer system. 
Another candidate for the tilted Dirac cone layer at the bottom can be organic material~\cite{SuzumuraReview} which has even smaller
Fermi velocity. 
In what follows we consider a double layer of borophene-graphene and study the effect of different Fermi velocity and chemical potential.
In this case the two branches of plasmons in the long wave length limit will be given by
\begin{align}
 & \omega_+^2=\frac{e^2 q\mu }{2}  \big(F^{(2)}(0)+G^{(1)}(\eta_1,\phi)\big),\nn\\
  &\omega_-^2= e^2 q^2 \mu d \frac{F^{(2)}(0)G^{(1)}(\eta_1,\phi) }{F^{(2)}(0)+G^{(1)}(\eta_1,\phi) },
 \label{collective-BG.eqn}
\end{align}
where the superscript in parenthesis indicates their layer indices. More explicitly, $F^{(2)}$ is the
same as function $F$ but specialized for layer $2$ whose Fermi velocity is $v_{F_2}$. The argument $0$ of
this function indicates that the tilt parameter $\eta_2=0$ as it stands for graphene layer. 
Similarly $G^{(1)}$ is the same function $G$ for the layer $1$ whose Fermi velocity is $v_{F_1}$, and its tilt is $\eta_1$. 

\begin{figure}[t]
 \includegraphics[width = .47\textwidth] {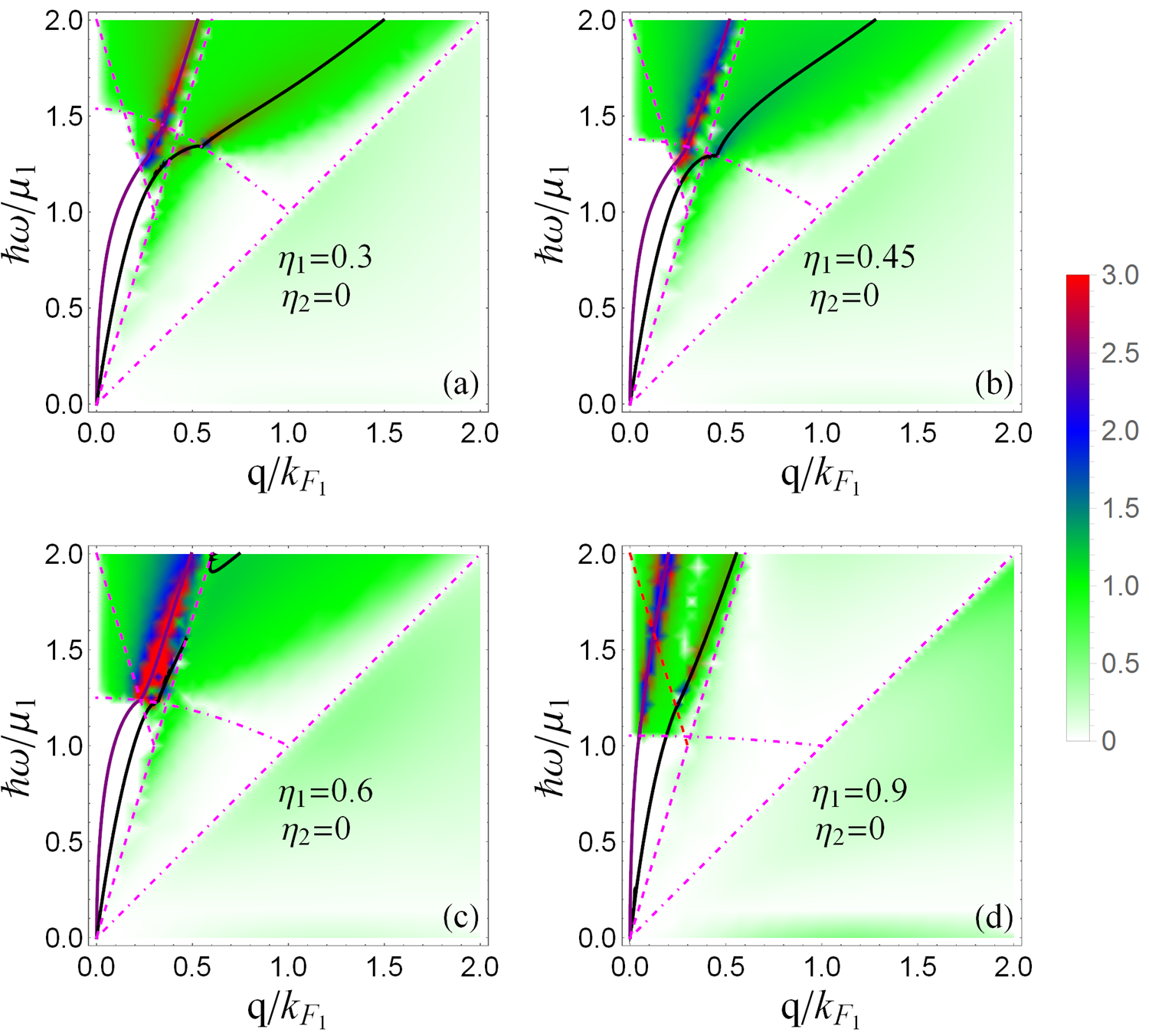}
  \caption{(Color online) Plasmon dispersions in DLBG system along with intensity plot of loss function $|\Im \varepsilon^{-1}(\bsq,\omega)|$.  
The direction of $\bsq$ is fixed by $\phi=\pi/2$, the layers separated by $k_{F_1}d=5.3$ have same chemical potential ($\mu_1=\mu_2$) and the tilting parameter of 
borophene layer (number $1$) in each panel is: (a) $\eta_1=0.3$, (b) $\eta_2=0.45$ (c) $\eta=0.6$, and (d) $\eta=0.9$. 
Purple and black solid lines are the plasmon dispersions for $\pm$ modes, respectively. 
The dotdashed (dashed) pink lines are the boundary of interband and intraband borophene (graphene) PHC. 
The vertical (horizontal) axis is the dimensionless energy $\hbar\omega/\mu_1$ (momentum $q/k_{F_1} $).
The graphene layer, $2$ has no tilting, $\eta_2=0$.}  
 \label{BG1.fig}
\end{figure}

First in Fig.~\ref{BG1.fig} we show the plasmon modes in the DLBG with equal chemical potential ($\mu_1=\mu_2$) but different Fermi velocity $v_{F_1}\neq v_{F_2}$.
As pointed out, the subscripts $1,2$ stand for borophene and graphene respectively. 
The tilting parameter for graphene layer, $\eta_2=0$ and tilting parameter for borophene layer in panel (a), (b), (c), (d) 
are taken to be $\eta_1=0.3, 0.45, 0.6 , 0.9$, respectively. The unit of energy is taken to be $\mu_1$ which equals $\mu_2$. 
However, since the Fermi velocities are different, for the unit of momentum one must specify either of the Fermi wave vectors, $k_{F_1}$ or $k_{F_2}$
as the unit of energy. We adopt the former, and therefore the horizontal axis is the dimensionless momentum $q/k_{F1}$,
and the vertical axis (as before) is the dimensionless energy $\omega/\mu_1$.
The color code is the same as previous figures. The
PHC boundary of borophene (graphene) has been defined by dotdashed (dashed) lines~\cite{Nishine2010,Tilted2018}.
As can be seen from, Fig.~\ref{BG1.fig} the PHC boundary of graphene has the larger slope as a result of its larger Fermi velocity value. 
Since the plasmons of monolayer of graphene are split off from its intraband PHC, in a combined DLBG system too, the level repulsion
from the intraband PHC of graphene pushes both modes to higher energies. This feature not only holds for the undamped portion of the 
plasmon branches, but it also holds for the damped portion of both branches that enters the interband PHC of the union of interlayer and
intralayer PH excitations. So the essential role of the difference in the velocity of the two layers is to sustain both
plasmon branches at velocities larger than the greater of the two. 

Note that in the DLBG system the PHC will be the union of intralayer PH excitations of both layers. 
In this way, the interband portion of the PHC for moderate $\eta_1$ comes below the $\omega_{\rm kink}$ curve. 
This is manifest in panels (a), (b) and (c) of Fig.~\ref{BG1.fig} where the dashed boundary (of graphene PHC)
has come below the dotdashed boundary (of the borophene PHC). In this way, the damping of $\pm$ modes in panels
(a) and (b) start at lower energy and momenta than anticipated from $\omega_{\rm kink}$ curve. 
Please note that, although the damping might start before the modes hit $\omega_{\rm kink}$ (dotdashed upper boundary),
but the kink always starts once the modes cross the $\omega_{\rm kink}$ boundary. 
This establishes that the $\omega_{\rm kink}$ very well deserves the subscript "kink".
Finally, again the generic property of both modes can be observe that
the energy of both modes increases by increasing the tilt parameter. 
Note that as argued for Fig.~\ref{eta0-eta.fig}, in the decoupled limit, only borophene layer
has kinks, while in the coupled graphene-borophene double layer, both dispersions 
have a kink at $\omega_{\rm kink}$.

\begin{figure}[t]
 \includegraphics[width = .49\textwidth] {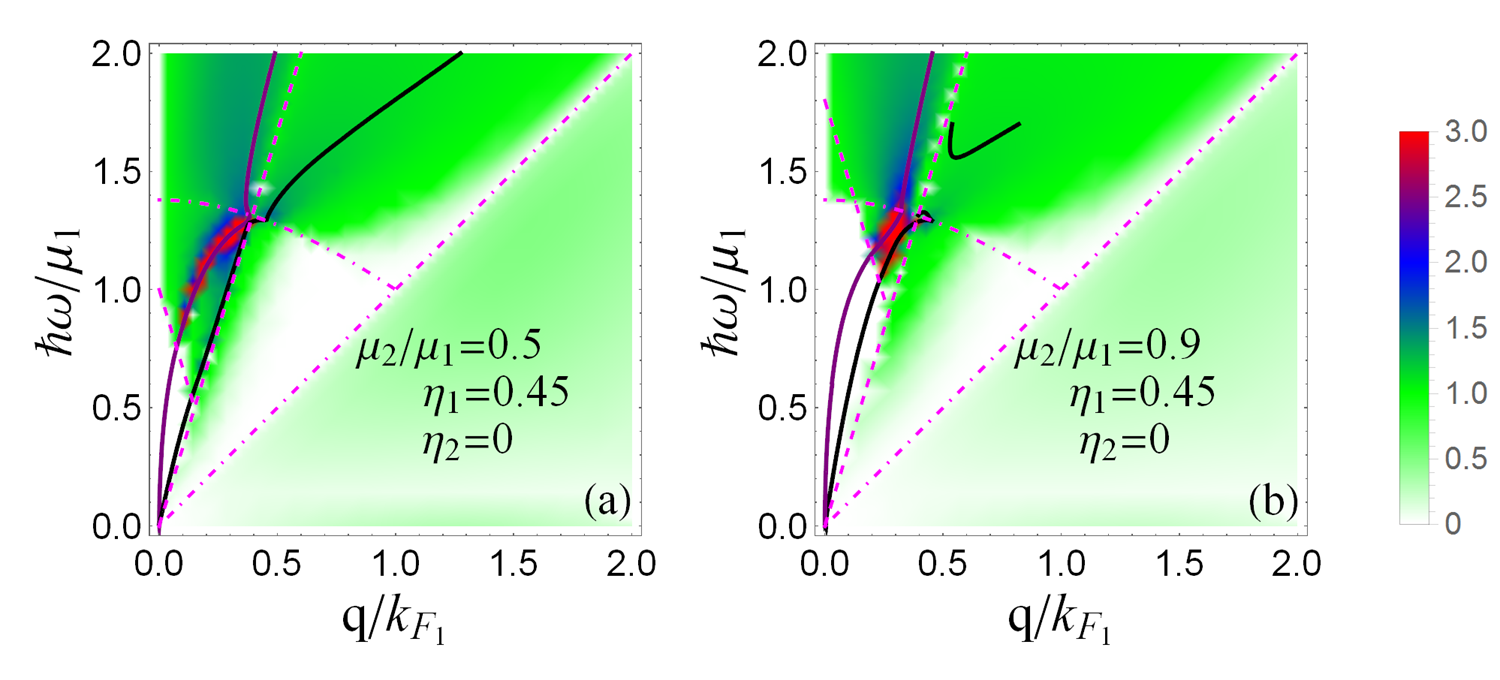}
   \caption{(Color online) Plasmon dispersions in DLBG for different chemical potential together with intensity plot of loss 
function $|\Im \varepsilon^{-1}(\bsq,\omega)|$. The direction of $\bsq$ is fixed by $\phi=\pi/2$ and the tilting parameter of borophene layer 
is $\eta_1=0.45$ which is separated by $k_{F_1}d=5.3$, where the chemical potential $\mu_1$ is fixed and is unit of energy. 
The chemical $\mu_2$ is given by the ratio $\mu_2/mu_1$ which is (a) $0.5$ and (b) $0.9$. 
The rest of the conventions are as in Fig.~\ref{BG1.fig}. 
}
 \label{BG2.fig}
\end{figure} 
Next, we consider DLBG with different chemical potential ($\mu_2\ne\mu_1$) and of course with different Fermi velocities in Fig.~\ref{BG2.fig}. 
In this figure the borophene layer 
is assumed to have the tilting $\eta_1=0.45$ and its chemical potential ($\mu_1$) is greater than the chemical potential of graphene 
($\mu_2$). As can be seen the undamped window for plasmon mode dispersion is more restricted as a result of different chemical potentials. 
Let us start by panel (b) where chemical potentials are different, but close to each other. In this case both in-phase and out-of-phase modes
are present, and their group velocity scale is set by the greater velocity (which belongs to graphene). Decreasing the chemical potential $\mu_2$ of
graphene, the "nearly" triangular window which is formed by the union of intralayer PHC of both layers, shrinks and the out-of-phase mode starts to 
sink into the intraband PHC dominated by PH excitaions of graphene. By further decrease in the $\mu_2$, the out-of-phase mode will entirely
disappear. This feature is similar to one considered in Fig.~\ref{BB1.fig}, where the out-of-phase mode is swallowed by the PHC.
In addition, as in all figures, both modes will have their kinks at their intersection with $\omega_{\rm kink}$.
Note that in panel (b) of Fig.~\ref{BG2.fig} and panel (c) of Fig.~\ref{BG1.fig} the out-of-phase mode is interrupted.
The region of interruption in both cases happens when the intrabanc PHC of graphene hits the plasmon branch. 
The density of PH excitations in intraband PHC are always much larger than the interband ones, and hence are
able to destroy the plasmon branches that hits this portion of PHC. 

\begin{figure}[t]
 \includegraphics[width = .47\textwidth] {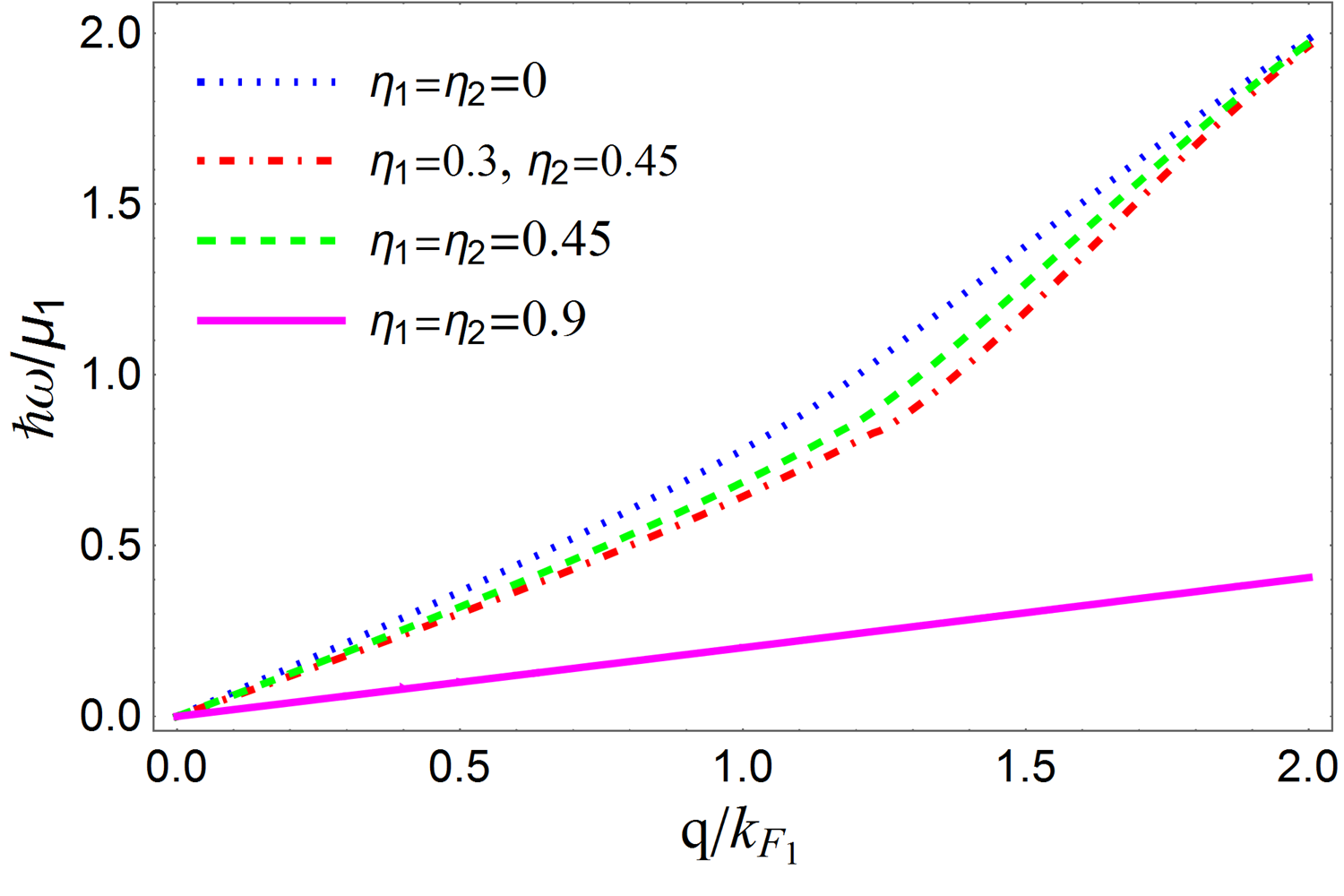}
   \caption{(Color online) Plasmon dispersions in DGL and DLB and with the same chemical potential. The direction of $\bsq$ is fixed by $\phi=\pi/2$. Here the tilting parameter for each line is: the blue dotted line points $\eta_1=\eta_2=0$ for DGL, the green dashed line points $\eta_1=0.3, \eta_2=0.45$ for DLB, the green dotdashed line points $\eta_1=\eta_2=0.45$ and the pink solid line points  $\eta_1=\eta_2=0.45$  for DLB. The vertical (horizontal) axis is the dimensionless quantity $\hbar \omega/\mu_1$ ($q/k_{F_1} $). The separation of layers is set by $k_{F_1}d=5.3$. } 
 \label{overdamped.fig}
\end{figure} 

\section{Overdamped Plasmon branch}\label{ODP}
In our previous work~\cite{Tilted2018}, we noted that an exclusive consequence of the tilt,
in addition to kinks in the dispersion of plasmons in monolayer system, is to provide
a unique chance for the emergence of an overdamped branch of plasmon excitations
which lies deep in the intraband PHC. Since the density of intraband PH excitations is
quite large, this provides a significant bath for Landau damping of this plasmon branch,
and therefore it gets quickly damped. Although this branch is heavily damped, 
but since it lives in lower energy than the standard plasmon branch, 
in time scales smaller than its lifetime $\tau$, it will be able to interact with
other low-energy excitations, including the single-particle excitations. Therefore it is
important to study this branch in the double layers as well. 

In Ref.~\onlinecite{Tilted2018}, we found that the overdamped plasmon branch for  borophene monolayer
is in the energy range $\omega<\omega_s$. This mode is 
caused by strong enough tilt, and disperses linearly. In the case of monolayer graphene where there
is no tilt, such an overdamped mode does not exist at all. It is interesting to note that when it comes to double layer
graphene, such an overdamped mode will appear. This has not been explored in earlier publications
addressing the double layer systems~\cite{Hwang2009,Gan2012,Profumo2012}. However, we find that even
for upright Dirac cones in a double layer system, an overdamped branch emerges. 
Fig.~\ref{overdamped.fig}, shows the dispersion of overdamped plasmon for a double layer composed of 
tilted Dirac cone systems, where Fermi velocities and chemical potentials and are the same.  
The dispersion has been plotted for $\phi=\pi/2$. The distance is fixed by $k_{F_1}d=5.3$.
Values of tilt parameter for each layer is indicated in the legend. 
As can be seen, even for $\eta_1=\eta_2=0$ there is an overdamped plasmon branch. 
The effect of tilt in each of layers is to reduce the energy of the overdamped plasmon
mode. The solid line represents the overdamped plasmon mode for quite large tilts $\eta_1=\eta_2=0.9$.
This mode disperses linearly over a much larger range of momenta, while for smaller values of
tilt parameters, the linear dispersion holds upto $q\sim k_{F_1}$. 
When one of the layers is doped and the other one is undoped, there will be no
overdamped solution.

\section{Summary and conclusion}\label{summary}
In this work, we investigated plasmon oscillations in double layer systems where
either one or both of the layers have tilted Dirac cone spectrum. It is well known that in this
context, there will be two plasmon modes. The in-phsae mode will disperse as $\sqrt q$ --
consistent with the hydrodynamic picture -- while
the out-of-phase mode disperses as $q^1$. 
The in-phase (symmetric) mode always lies at higher energies that the out-of-phase (asymmetric) mode. 
This is in contrast to the intuition from molecular orbitals where the symmetric combination of atomic 
orbitals usually has lower energy than the asymmetric combination. This is because in the present case,
we are dealing with a symmetric combination of particle-hole objects, and not single-particle orbitals.
An extra minus sign coming from the fermion loop places the symmetric plasmons at higher energies.

When the tilted Dirac cone systems are
combined in a double layer framework, interesting plasmonic features arises. 
The tilt of the Dirac cone
is manifested in its plasmons as a kink when it crosses $\omega_{\rm kink}$. Such a
kink is absent in Dirac cone without tilt. In a bilayer setting we find quite generically
that even when one of the layers hosts tilted Dirac cone, the plasmonic kink will be
inherited by both in-phase and out-of-phase mode. This kink in both branches takes place
at precisely $\omega_{\rm kink}$. In situation such as in Fig.~\ref{BB2.fig} where due to difference
in the chemical potential of the two tilted Dirac cone layers there are two $\omega_{\rm kink}$ energy 
scales, each of the plasmon branches develops a kink upon crossing every $\omega_{\rm kink}$ (dotdashed pink) curve. 
In this situations there will be a total number of four kinks; two kinks for every plasmon branch. 

When one of the layers is graphene with larger Fermi velocity, the small window where
undamped plasmons can live will become smaller and will be set by the larger Fermi velocity 
of graphene. This pushes both in-phase and out-of-phase plasmon modes to higher energy.
Therefore the typical plasmonic group velocity in such double layer systems with two 
different Fermi velocities, is set by the greater of the two velocities. 

Another unique feature of tilted Dirac cone monolayer is the existence of
linearly dispersing overdamped plasmon mode inside the intraband PHC. 
Although this mode does not exist in monolayers of upright Dirac cone systems
such as graphene, in the double layer setting such a mode emerges. 
In the double layer systems with tilt, this mode continuously reduces its
slope by increasing the tilt. This mode is the in-phase overdampled oscillation
of the individual tilted layers~\cite{Tilted2018}. 

%\section{Acknowledgements}
%SAJ thanks R. Fazio for invitation to the Abdus Salam international center for theoretical
%physics where this work was completed. 

\bibliography{DL}% Produces the bibliography via BibTeX.

\end{document}